\documentstyle[amssymb,preprint,aps,floats,psfig,prb,tighten]{revtex}
\newcommand{\Va}{$\alpha$'- NaV$_2$O$_5\;$}

\begin{document} \draft \title{Magnetic excitations in charge ordered \Va}
\author{P. Thalmeier$^1$ and A. N. Yaresko$^2$}
\address{$^1$Max-Planck-Institute for Chemical Physics of Solids\\
$^2$Max-Planck-Institute for the Physics of Complex Systems\\ D-01187 Dresden,
Germany} \date{\today} \maketitle

\begin{abstract} An investigation of the spin excitation spectrum of charge
ordered (CO) \Va is presented. We discuss several different exchange models
which may be relevant for this compound, namely in- line and zig-zag chain
models with weak as well as strong inter- chain coupling and also a ladder
model and a CO/MV (mixed valent) model. We put special emphasis on the
importance of large additional exchange across the diagonals of V- ladders and
the presence of exchange anisotropies on the excitation spectrum. It is shown
that the observed splitting of transverse dispersion branches may both be
interpreted as anisotropy effect as well as acoustic- optic mode splitting in
the weakly coupled chain models. In addition we calculate the field dependence
of excitation modes in these models. Furthermore we show that for strong inter-
chain coupling, as suggested by recent LDA+U results, an additional high energy
optical excitation appears and the spin gap is determined by anisotropies.
The most promising CO/MV model predicts a spin wave dispersion perpendicular to
the chains which agrees very well with recent results obtained by inelastic
neutron scattering.  \end{abstract}


\section{Introduction} Transition metal- oxygen pyramids are ideal building
blocks to obtain insulators with low D structures of 3d- ions like chains or
ladders. Their localized spins exhibit collective quantum properties at low
temperatures, e.g. spin gap formation in S=1 chains or S=1/2 ladders. In
addition there is the possibility of spin gap appearance due to the spin-
Peierls (SP) mechanism which causes dimerization of the chain. The standard
example is CuGeO$_3$ \cite{Hase}. Recently \Va which has the Trellis lattice
structure with alternating shifted ladders (Figs.1 and 6) was investigated for
similar reasons: observation of a superstructure below T$_c$= 33 K and
subsequent spin gap formation as witnessed by a drop of the susceptibility
below T$_c$ \cite{Isobe}. However it is clear now that this compound does not
exhibit a standard SP- transition because above T$_c$ it is a homogeneous mixed
valent (MV) insulator with one 3d- electron per V-V rung. Therefore above T$_c$
\Va is a quarter filled ladder system with equivalent V- sites instead of a
family of half filled (atomic spin) chains. This was concluded from x- ray
\cite{Chatt1,Chatt2,Smol} and NMR- experiments \cite{Ohama}. They also show
that below T$_c$ in the dimerized state two inequivalent V- sites exist.
Therefore a charge ordering (CO) transition which localizes the V 3d- electrons
on one site of each rung of the ladders must take place at T$_c$. Possible CO-
structures have been discussed by various authors \cite{Chatt2,Thal,Seo} but so
far the real low temperature structure remains controversial. In general a CO
transition may occur when the {\em inter}- site Coulomb repulsion is larger
than kinetic energy terms, this is only possible in low carrier density
semimetals or insulators like \Va. Charge ordering can be viewed as a Wigner-
crystallization on a lattice \cite{Fulde}. This should not be confused with the
CDW transition in more metallic systems. The CO- mechanism in insulating \Va
can be described within an effective frustraded 2D- Ising model\cite{Thal}. It
leads to in-line or zig-zag charge order depending on whether the difference in
Coulomb repulsion, K$_1$-K$_2$ between n.n. ( K$_1$) and n.n.n. (K$_2$) is
positive or negative respectively. Later we will also discuss alternative CO
structures. In Ref.~\onlinecite{Thal} the possible origin of spin gap formation
was discussed for the in-line structure where an induced SP transition slightly
below the primary in-line CO transition was proposed. This scenario would
naturally explain the appearance of two superposed phase transitions from
thermal expansion measurements \cite{Geibel} and the observed anomalous BCS-
ratio. As mentioned the zig-zag CO is an alternative possibility, it has been
discussed in Ref.~\onlinecite{Seo} and a related structure in
Ref.~\onlinecite{Chatt2}. It has been claimed, though not discussed in any
detail that this structure leads directly to a gap in spin-
excitations.\\[0.5cm] Important information on the true low temperature CO
structure may be obtained from an investigation of the complete dispersion of
magnetic excitations, especially along $\vec{a}^*$ ($\perp$ to the chain axis
$\vec{b}^*$). However the existing neutron scattering results \cite{Yosi} were
rather limited in resolution. A special behaviour of excitations for wave
vector $\vec{q}$=(q$_x$,$\pi$)(in units of $\frac{1}{a}$ and $\frac{1}{b}$) was
proposed: The spin gap mode with $\Delta_s$= 10 meV was suggested to be twofold
degenerate at q$_x$= 0, 2$\pi$ and to split into two excitations about 2-3 meV
apart for intermediate q$_x$. This was also discussed in a theoretical
model\cite{Val}. But more recent experiments with much better resolution
\cite{Regnault} have shown that this is definitely not true and a splitting of
$\sim$1 meV exists even at the points $\vec{q}$= (0$,\pi$) and (2$\pi$,$\pi$).
Furthermore new electronic structure calculations \cite{Yar} based on the LDA+U
approach have shown that there is an additional important exchange coupling
which has previously been neglected. In addition like in the cuprates small
exchange anisotropies may also lead to gaps for spin excitations. Therefore it
is desirable to develop a general theory of magnetic excitations in \Va that
incorporates all these aspects and allows to calculate all possible features of
the spin excitations in the various candidate CO- structures of \Va, including
the effect of an external field.  \\[0.5cm] In the following the exchange model
for the CO- structures is defined (section II). For the low temperature CO
phases with intra-chain dimerization it may be mapped to a simplified model
including only relevant dimer variables (section III). In section IV the spin
dynamics of various exchange models for \Va will be investigated including
exchange anisotropies and external field. The resulting collective magnetic
excitations are studied for all models under special emphasis of the importance
of intra-chain exchange anisotropies and their influence on the mode
dispersions perpendicular to the chain ($\vec{b}$-) axis. Finally our
calculations and their connection to experimental results are summarized in
section V.

\section{Electronic structure, charge order and exchange models} In the high
temperature phase (T$>$T$_c$) \Va is an insulating mixed valence compound whose
electronic structure is now reasonably well understood \cite{Smol,Yar}. In an
effective tight binding (TB) model including only V(3d) orbitals one has
bonding (B) and antibonding (AB) bands corresponding to the symmetric and
antisymmetric molecular orbitals of each V-V rung. In the following a similar
convention for the notation of TB hopping matrix elements is used as for the
exchange integrals in Figs. 1,6. The B-AB gap $\sim\tilde{t}$ is about one eV
and the band widths are $\sim$ 0.5 eV (B) and almost zero (AB). This difference
has an important origin \cite{Yar} which was not realized previously: Since the
dispersion of B and AB bands are proportional to t+t$_d$ and t-t$_d$
respectively it means that t$_d$$\simeq$ t, and hence t$_d$, the hopping across
the ladder diagonal cannot be neglected and is necessary for a realistic TB
model of {\em both} B and AB bands. In a naive superexchange model this would
also mean that the
 AF exchange constants J$\sim\frac{2t^2}{U}$ and J$_d\sim\frac{2t_d^2}{U}$
 should be of the same order of magnitude. This is indeed confirmed by
 spin-polarized LDA+U calculations \cite{Yar} where CO for the 3d- electrons in
 the V-V rungs has to be assumed. They also show that CO \Va is in an
 insulating state for sufficiently large on-site U$\geq$ 3 eV contrary to
 conventional LDA-calculations which predicts a metallic state. As a mean field
 like theory with broken orbital symmetry the LDA+U approach does of course not
 describe the true microscopic nature of the disordered MV insulating state
 above T$_c$. This is still an open problem. The transition from the high
 temperature MV state to the CO state was investigated in
 Ref.~\onlinecite{Thal}. It was described within a frustrated 2D Ising model
 where the Ising spin $\tau_z$=$\pm$1 denotes the 3d- electron localized on the
 right or left position of the rung. In this context the CO of Fig.1a and
 Fig.1b can then be described by an order parameter
 $\langle\tau_z\rangle_{\vec{Q}}=\sum_i\langle\tau^i_z\rangle\exp(i\vec{Q}\vec{R}_i)$.
 For $\vec{Q}$=0 one obtains the "ferro-" type in- line CO structure and for
 $\vec{Q}$=($\pi$,$\pi$) the "antiferro-" type  zig-zag CO structure depending
 whether K(0)=K$_1$-K$_2$$>$0 or K($\vec{Q}$)=K$_2$-K$_1$$>$0. In this way
 model Hamiltonians of the effective Ising type as in Ref.~\onlinecite{Thal} or
 extended Hubbard models in Hartree-Fock approximation \cite{Seo} may be used
 to describe the CO transition in a qualitative way. However it is illusory to
 use such model Hamiltonians in an attempt to actually predict the most
 favorable CO structure. This requires a method like LDA+U which can provide
 ab-initio (aside from U) total energies of the various CO structures. It has
 been sucessfully used for CO- phenomena in semimetallic 4f- compounds
 \cite{Ant} and may also be a powerful method for the vanadates \cite{Yar}. In
 this work the CO mechanism itself is not considered. We rather start from
 plausible candidate structures at low temperature and an appropriate exchange
 model. The derivation of an exchange model  for \Va from an original extended
 Hubbard model was described in Ref.~\onlinecite{Thal} and is briefly
 recapitulated here. It proceeds by eliminating high energy charge fluctuations
 between the rungs, thus confining one d-electron or spin  within each rung.
 Within this subspace the original model may be mapped to an effective low
 energy Hamiltonian containing the d-electron spin and (Ising) pseudo- spin
 degrees of freedom, the latter describes which of the degenerate V- positions
 in the rung is occuppied. This Hamiltonian is formally similar to those used
 for the manganites where the pseudospin describes an orbital degeneracy of Mn
 ions. The Ising variable describes the CO transition and for T$\ll$ T$_c$
 where the intra- rung charge fluctations are also frozen we may replace it by
 its expectation value, i.e. the CO parameter. In this way the effective
 Hamiltonian reduces to an effective spin exchange Hamiltonian only, however
 with an exchange constant J$_{nm}$ (n,m= V-sites) that {\em depends} on the CO
 parameter, i.e. on the degree of charge disproportionation between the
 inequivalent V-atoms in \Va. In this low temperature approximation which we
 use here the actual size of the CO parameter is absorbed in the exchange
 constants and influences {\em only} the energy scale of the spin dynamics, the
 form of the exchange Hamiltonian (T$\ll$ T$_c$) is of the usual type as for
 spins in a completely CO system. In our case due to the orthorhombic symmetry
 it is essential to include exchange anisotropies which may be important for
 small mode splittings as observed in \Va. The model exchange Hamiltonian for
 the proposed CO structures in Figs.1,6 is then given by

\begin{eqnarray} H&=&\frac{1}{2}\sum_{n,m}(J^x_{nm}S^x_nS^x_m
+J^y_{nm}S^y_nS^y_m +J^z_{nm}S^z_nS^z_m)\nonumber\\ &&-g\mu_BH\sum_nS^z_n
\label{Ham} \end{eqnarray}

Here J$^\alpha_{nm}$ ($\alpha$ =x,y,z) denotes both inter- and intra- chain
couplings which may be different along the three crystal axis $\vec{a},
\vec{b}, \vec{c}$ (x,y,z). Note that part of the anisotropy in Eq.(\ref{Ham})
may be due to a Dzyaloshinski-Moriya interaction which can be transformed away
in 1D in a manner described in Ref.~\onlinecite{Oshi} and references cited
therein. A Zeeman term with field direction perpendicular to the Vanadium ab-
planes is also included to study the field dependence of excitations. Which
exchange couplings have to be used depends on the CO structure, i.e. the
position of the V$^{4+}$ S=$\frac{1}{2}$ spins because exchange bonds to
V$^{5+}$- ions with no d-electrons and S=0 are irrelevant for the spin
dynamics. This is shown in Figs.1 and 6 with sets of intra- (J, J$_d$) and
inter- chain (J', J'$_d$, J$_l$) exchange parameters (the cartesian index
$\alpha$ is suppressed). The former may be dimerized to
J$_{1,2}$=J(1$\pm\delta$) (Fig.1a) and J$_{1,2}$=J$_d$(1$\pm\delta$) (Fig.1b).
This set has been enlarged as compared to Ref.~\onlinecite{Thal} where only J,
J' were included. Note that J$_d$ and J' are not contributing in the in-line CO
structure of Fig.1a; J is inactive for the zig-zag structure (Fig.1b) and J' is
not relevant in the structures of Fig.6. In this work we consider the following
cases: (1) quasi-1D models either in the in-line, zig-zag or ladder CO where
the inter- chain or -ladder couplings J',J'$_d$ etc. are assumed to be much
smaller than the intra- chain couplings J,J$_d$ or the intra- ladder
$\tilde{J}$. (2) a quasi- 2D model where J' is of the same order as J and
J$_d$. This possibility has been suggested by recent LDA+U results. (3) a mixed
CO/MV structure which will be discussed later. Different methods have to be
used for calculating the excitation spectrum in these cases. Figs.1a,b show
that J and J$_d$ play the same role for in-line and zig-zag quasi- 1D models
respectively. Therefore one has in both cases quasi- 1D spin chains with intra-
chain coupling J (in-line) or J$_d$ (zig-zag) coupled by small inter- chain
interactions J'$_d$ (in-line) and J' (zig-zag). There is one essential
difference however: In the in-line structure the CO- transition itself does not
lead to a dimerization of the chain with an intra-chain J along $\vec{b}$. This
may be due to a secondary SP- transition slightly below \cite{Thal} leading to
a dimerization J $\rightarrow$ J(1$\pm \delta$) with $\delta \ll$ 1. On the
other hand CO in the zig-zag structure may itself be accompanied by a lattice
distortion such that the two legs of the zig- zag chain have different length
leading directly to a dimerized exchange J$_d$(1$\pm\delta$). However it is
possible that even in this structure the most important contribution to the
dimerization comes from the exchange energy J$_d$ along the zig-zag legs in
Fig.1b. Irrespective of the origin of dimerization a spin gap opens for both CO
chain structures with its size $\Delta_s(\delta)$ depending on dimerization
strength. On the other hand if a spin ladder structure as Fig.6a is realised in
\Va a spin gap will appear already without dimerization along $\vec{b}$.
Finally in the quasi- 2D model with strongly coupled chains and in the mixed
CO/MV model a broken symmetry spin wave (SW) calculation will show that the
spin gap can be attributed to pure anisotropy effects.\\ For the single
dimerized chain or the spin ladder methods based on the Jordan-Wigner
transformation \cite{Uhrig,Azzouz} exist to investigate the excitation
spectrum. However the focus in this work is primarily on the typical behaviour
of the {\em transverse} dispersion ($\vec{q}\perp\vec{b}^*$) of excitations
where the influence of interchain coupling and exchange anisotropies has to be
studied. For this purpose it is necessary to use a simple theory as starting
point for the intra- chain excitations ($\parallel \vec{b}^*$). It is
physically appealing to use a spin dimer representation where the presence of a
spin gap is already manifest in the local dimer basis as a singlet-"triplet"
splitting. This representation may also be mapped to the so called bond boson
model introduced in Ref.~\onlinecite{Gopa} for spin ladders. However, for the
purpose of investigating spin excitations only it is more convenient to keep
the original spin- dimer basis, especially when the effect of exchange
anisotropies on excitations is to be considered. The basic features of the
spin- dimer representation following Ref.~\onlinecite{Leue} are outlined in the
next section and adapted to the relevant CO spin structures on the Trellis
lattice.

\section{The local spin dimer model} In the dimerized phase of the chain models
Fig.1a,b or in the case of a ladder with $\tilde{J}$ $>$ J (Fig.6a) it is a
useful approach to start from a basis where the strongest exchange pairs i.e.
J(1+$\delta$) or $\tilde{J}$ respectively are diagonalized exactly and the
weaker couplings are treated perturbatively in random phase approximation
(RPA). This method has first been used in Ref.~\onlinecite{Leue} in a different
context. Presently this means the introduction of dimer variables

\begin{eqnarray} \vec{K}_i&=&\vec{S}_{1i}+\vec{S}_{2i} \nonumber\\
\vec{L}_i&=&\vec{S}_{1i}-\vec{S}_{2i} \end{eqnarray}

for each pair of strongly coupled dimer spins ($\vec{S}_{1i},\vec{S}_{2i}$)
where $\vec{R}_i$ denotes the positions in the dimer covering lattice. Using
this mapping the Hamiltonian in Eq.(\ref{Ham}) may be transformed to

\begin{eqnarray} H&=& \frac{1}{4}\sum_{i\alpha}J_1^\alpha(K_i^\alpha K_i^\alpha
- L_i^\alpha L_i^\alpha ) -g\mu_BH\sum_iK^z_i\nonumber\\
&&-\frac{1}{8}\sum_{\langle ij\rangle\alpha}J_2^\alpha L_i^\alpha L_j^\alpha
-\frac{1}{8}\sum_{\ll ij\gg\alpha}J_3^\alpha(i,j)L_i^\alpha L_j^\alpha
\label{DHam} \end{eqnarray}

Here the first and second term $\sim$ J$_1^\alpha$ describes the local dimer
energy and the Zeeman energy respectively, the third term $\sim$ J$_2^\alpha$
denotes the n.n. dimer interactions along the chain direction  $\vec{b}$ and
the last term  J$_3^\alpha$ interactions of dimers on different chains. For the
two chain CO models we have $J_1^\alpha$ = J$_\alpha (1+\delta)$, J$_2^\alpha$
= J$_\alpha (1-\delta)$ (with J$_\alpha >$0 for AF intra- chain exchange) and
J$_3^\alpha$ depends on the specific model discussed. Since J and J$_d$ play
the same role in the in-line and zig-zag model respectively we formally
identify J$_d$$\rightarrow$ J in subsequent discussions of these two models.
For the ladder model one has J$_1^\alpha$ =$\tilde{J}_\alpha$ and
J$_2^\alpha$$\equiv$ J$_e^\alpha$= J$_d^\alpha$-J$^\alpha$. In the Hamiltonian
of Eq.(\ref{DHam}) irrelevant parts containing terms $\sim$ K$^\alpha _i$
K$^\alpha _j$ and  L$^\alpha _i$ K$^\alpha _j$ are not included because they do
not have matrix elements from the singlet ground state to the excited states
and hence do not contribute to the dispersion of spin excitations \cite{Leue}.
The energies and states of the S=$\frac{1}{2}$ dimer are given by

\begin{eqnarray} E_1&=&0 \nonumber\\
E_2&=&J_1=\frac{1}{2}(J_1^x+J_1^y)\nonumber\\
E_3&=&J_1-j'_1+j_1=\frac{1}{2}(J_1^x+J_1^z)=\Delta'\nonumber\\
E_4&=&J_1-j'_1=\frac{1}{2}(J_1^y+J_1^z)=\Delta\nonumber\\
|\psi_1\rangle&=&\frac{1}{\sqrt{2}} (|\uparrow\downarrow\rangle
-|\downarrow\uparrow\rangle)\label{States}\\
|\psi_2\rangle&=&\frac{1}{\sqrt{2}} (|\uparrow\downarrow\rangle
+|\downarrow\uparrow\rangle)\nonumber\\ |\psi_3\rangle&=&\frac{1}{\sqrt{2}}
(|\uparrow\uparrow\rangle +|\downarrow\downarrow\rangle)\nonumber\\
|\psi_4\rangle&=&\frac{1}{\sqrt{2}} (|\uparrow\uparrow\rangle
-|\downarrow\downarrow\rangle)\nonumber \end{eqnarray}

The ground state singlet $|\psi_1\rangle$ is separated by an energy $\sim$
J$_1$ from the triplet states which are slightly split by an energies
j'$_1$-j$_1$ and j'$_1$ due to the exchange anisotropies given by
j$_1$=$\frac{1}{2}$(J$_1^x$-J$_1^y$), j'$_1$=$\frac{1}{2}$(J$_1^x$-J$_1^z$)
where $|j_1|$, $|j'_1|\ll$ $|J_1|$. In the isotropic case $j_1$= $j'_1\equiv$ 0
the excited states form a dimer triplet at $\Delta=\Delta'\equiv J_1$. The
dipolar matrix elements
$|M_\alpha^i|^2$=$|\langle\psi_1|L_\alpha|\psi_i\rangle|^2$ calculated from
Eq.(\ref{States}) are given by $|M^3_y|^2$= $|M^4_x|^2$=1 and zero else.
Therefore in the dynamical spin susceptibility u$_{\alpha\beta}(\omega)$ of a
single dimer only E$_3 =\Delta'$ and E$_4 =\Delta$ appear as possible dimer
excitations which is obvious from the form of dimer states $|\psi_i\rangle$ in
Eq.(\ref{States}). For the present zero field case u$_{\alpha\beta}(\omega)$=
u$_{\alpha\alpha}(\omega)\delta_{\alpha\beta}$ ($\alpha, \beta$= x,y) with

\begin{equation} u_{xx}(\omega)=\frac{2\Delta}{\Delta^2-\omega^2}\;\;\;\;\;\;
u_{yy}(\omega)=\frac{2\Delta'}{\Delta'^2-\omega^2}\label{Single} \end{equation}

Due to both intra- and inter- chain interactions the two local dimer
excitations at $\Delta, \Delta'$ will turn into dispersive propagating modes
whose minimum energy is the spin gap $\Delta_s$. Before we discuss this in
detail in the next section we first investigate the effect of an external field
$\parallel$ c on the dimer states described by the Zeeman term in
Eq.(\ref{DHam}). Because K$_z$ commutes with $\vec{K}^2$ one has
$\langle\psi_1|K_z|\psi_i\rangle$=0; i.e. no mixing of singlet and triplet
states. The only non zero matrix element is $\langle\psi_3|K_z|\psi_4\rangle$,
therefore the energies E$_{1,2}$ and states   $|\psi_{1,2}\rangle$ will be
unchanged but E$_{3,4}$ and $|\psi_{3,4}\rangle$ become field dependent:

\begin{eqnarray} |\psi_+\rangle&=& u|\psi_3\rangle +v|\psi_4\rangle\nonumber\\
|\psi_-\rangle&=&-v|\psi_3\rangle +u|\psi_4\rangle\nonumber\\
u^2&=&\frac{h^2}{h^2+[(j_1^2+h^2)^\frac{1}{2}-j_1]^2}\\
v^2&=&\frac{[(j_1^2+h^2)^\frac{1}{2}-j_1]^2 }
{h^2+[(j_1^2+h^2)^\frac{1}{2}-j_1]^2}\label{Trans}\nonumber \end{eqnarray}

The energies of the new eigenstates $|\psi_\pm\rangle$ are given by

\begin{eqnarray}
E_+&=&\Delta'(h)=J_1-j'_1+\frac{1}{2}j_1[1+(1+(\frac{h}{j_1})^2)^\frac{1}{2}]
\nonumber\\ E_-&=&\Delta(h)
=J_1-j'_1+\frac{1}{2}j_1[1-(1-(\frac{h}{j_1})^2)^\frac{1}{2}] \label{Delta}
\end{eqnarray}

In the limit h $\rightarrow$ 0 $|\psi_\pm\rangle$ $\rightarrow$
$|\psi_{3,4}\rangle$ and E$_\pm$ $\rightarrow$ E$_{3,4}$. The local dimer
susceptibility u$_{\alpha\beta}(\omega)$ is now given by

\begin{eqnarray} u_{xx}(\omega)&=&\frac{2u^2\Delta(h)}{\Delta(h)^2-\omega^2}
+\frac{2v^2\Delta'(h)}{\Delta'(h)^2-\omega^2}\nonumber\\
u_{yy}(\omega)&=&\frac{2u^2\Delta'(h)}{\Delta'(h)^2-\omega^2}
+\frac{2v^2\Delta(h)}{\Delta(h)^2-\omega^2}\label{HSingle}\\
u_{yx}(\omega)=-u_{xy}(\omega)&=&2i\omega uv\frac{\Delta(h)^2-\Delta'(h)^2}
{(\Delta(h)^2-\omega^2)(\Delta'(h)^2-\omega^2)}\nonumber \end{eqnarray}

The nondiagonal part is induced by the field. Eq.(\ref{HSingle}) fully
describes the local dimer magnetic response and is the basis for the
determination of the collective excitations in the various CO structures of the
Trellis lattice (Fig.1 and 6b).

\section{Collective magnetic excitations}

In the previous section the effect of the largest intra- dimer exchange
interaction J$_1^\alpha$ has been treated exactly within the single dimer
subspace. The effect of inter- dimer exchange may now be treated perturbatively
within random phase approximation (RPA). In this method the collective magnetic
excitations of the chain or ladder system are given by the dynamical RPA
susceptibility

\begin{equation} \tensor{\chi}(\vec{q},\omega)=
[\tensor{1}-\tensor{J}(\vec{q})\tensor{u}(\omega)]^{-1}\tensor{u}(\omega)
\equiv\tensor{D}^{-1}(\vec{q},\omega)\tensor{u}(\omega)\label{RPA}
\end{equation}

Here $\tensor{u}(\omega$) is the local dimer susceptibility tensor of
Eq.(\ref{HSingle}) and $\tensor{J}(\vec{q})$ the exchange tensor between the
{\em dimers} which depends on the specific CO- model considered,
$\vec{q}$=(q$_x$,q$_y$) is a wave vector in the reciprocal a$^*$b$^*$- plane in
units of $\frac{1}{a}$ and $\frac{1}{b}$. The tensors in Eq.(\ref{RPA}) have
double indices: Cartesian $\alpha,\beta =x,y,z$ as well as CO- sublattice
$\lambda, \tau$ =A,B. Explicitly
J$_{\lambda\tau}^{\alpha\beta}$($\vec{q}$)= 
$\delta_{\alpha\beta}$J$_{\lambda\tau}^{\alpha\alpha}$($\vec{q}$)and
u$_{\lambda\tau}^{\alpha\beta}(\omega)$=$\delta_{\lambda\tau}$u$_{\alpha\beta}$($\omega$).
For two sublattice CO- structures and two local dimer excitations
$\Delta,\Delta'$ with x,y polarisation one has to expect four ($\kappa$=1-4)
collective excitation branches $\omega_\kappa(\vec{q})$. They are given as
poles of $\tensor{\chi}(\vec{q},\omega)$ or zeroes of D($\vec{q},\omega$).
Strictly speaking this treatment is only valid when the intra- dimer exchange
is appreciably  larger than the inter- dimer coupling. For example in the
dimerized chain models of Fig.1a,b the limit $\delta$$\rightarrow$ 0 is
problematic because then J$_1$ $\rightarrow$ J$_2$ i.e. intra- and inter- dimer
exchange become equal. As shown below, Eq.(\ref{RPA}) nevertheless leads to the
qualitatively correct behaviour for the spin gap $\Delta_s(\delta\rightarrow 0)
\rightarrow 0$ although with a different scaling exponent. This indicates that
the present approach is more effective than the bond- boson theory in MF-
approximation \cite{Gopa} which leads to a singular $\Delta_s$ for $\delta$=
0.

\subsection{Excitations for single dimerized chains} To separate the effects of
intra- chain exchange anisotropies from those of inter- chain or sublattice
coupling it is useful to analyse first the single chain case at zero field.
Then J$_{\lambda\tau}^{\alpha\alpha}$($\vec{q}$)=
J$_D^{\alpha\alpha}$($\vec{q}$)$\delta_{\lambda\tau}$ is diagonal in the dimer
sublattice basis and Eq.(\ref{RPA}) factorizes for x,y polarisation and only
two modes exist. The resulting zeroes of $\tensor{D}_{x,y}(\vec{q},\omega)$ are
then the two propagating dimer excitations $\omega_{x,y}(\vec{q})$ where
$\vec{q}=q\vec{b}^*$ is directed along the chain direction. The result applies
both for the single linear chain and the zig-zag chain in Fig.1a,b (with
renaming J$_d$ $\rightarrow$ J implied as explained in the previous section).
Using Eq.(\ref{Single}) and the appropriate J($\vec{q}$) the mode dispersions 
are obtained as

\begin{eqnarray} \omega_x^2(q)&=&\frac{1}{2}(J_y+J_z)
[\frac{1}{2}(J_y+J_z)(1+\delta)^2\nonumber\\ &&-J_x(1-\delta^2)cos2q]\\
\omega_y^2(q)&=&\frac{1}{2}(J_x+J_z)
[\frac{1}{2}(J_x+J_z)(1+\delta)^2\nonumber\\
&&-J_y(1-\delta^2)cos2q]\label{Anis}\nonumber \end{eqnarray}

The spin gap $\Delta_s(\delta)$ is obtained as the minimum of
$\omega_{x,y}(q)$. The x,y- mode splitting at the q=0 is then given by

\begin{eqnarray}
\omega_x^2(0)-\omega_y^2(0)&=&\frac{1}{2}(J_x-J_y)(1+\delta)\nonumber\\
&&[(1+\delta)(\Delta+\Delta')-(1-\delta)J_z] \end{eqnarray}

which is proportional to the in-plane anisotropy j$_1$=
$\frac{1}{2}$(J$_x$-J$_y$)(1+$\delta$). If j$_1\equiv$ 0 the x,y modes are
degenerate which can already be seen from their corresponding local dimer
excitations E$_3$, E$_4$ in Eq.(\ref{States}). For $\delta$= 0 one has

\begin{eqnarray} \omega_x=(\Delta D_x)^\frac{1}{2}
\;\;\;;D_x=\frac{1}{2}[J_y+J_z-2J_x]>0 \nonumber\\ \omega_y=(\Delta
D_y)^\frac{1}{2} \;\;\;;D_y=\frac{1}{2}[J_x+J_z-2J_y]>0 \end{eqnarray}

For the uniaxial case, using J$_z>$J$_x$=J$_y$ without loss of generality and
D$_{x,y}$=$\frac{1}{2}$[J$_z$-J$_{x,y}$] this leads to
$\omega_{x,y}=\frac{1}{2}(J_z^2-J_{x,y}^2)^\frac{1}{2}\equiv\Delta_s$. In this
limit the spin gap is a pure anisotropy gap. Approaching the Heisenberg case
$\Delta_s$ vanishes. It is also instructive to consider the dispersion
Eq.(\ref{Anis}) directly for the Heisenberg case for $\delta\geq 0$:

\begin{eqnarray}
\omega_{x,y}^2\equiv\omega^2(q)=2J^2(1+\delta)(\sin^2q+\delta\cos^2q)
 \label{Iso} \end{eqnarray}

This leads to a spin gap given by

\begin{equation} \frac{\Delta_s}{J}=(2\delta)^\frac{1}{2}(1+\delta)^\frac{1}{2}
\end{equation}

For $\delta\rightarrow$ 0 it vanishes like $\Delta_s\sim\delta^\frac{1}{2}$.
Thus the spin dimer RPA approximation gives again a qualitatively correct
behaviour although the $\delta$- scaling exponent $\frac{1}{2}$ is smaller than
the exact one \cite{Uhrig} which is $\frac{2}{3}$. The dispersion
Eq.(\ref{Iso}) reduces to

\begin{equation} \omega(q)=\alpha J\sin q \label{DCP} \end{equation}

for the undimerized chain whith $\alpha=\sqrt{2}$. This is slightly smaller
than the value $\alpha_{DCP}=\pi/2$ for which Eq.(\ref{DCP}) describes the
lower boundary of the exact Des Cloizeaux Pearson (DCP) excitation spectrum of
the 1D HAF \cite{DCPP}. Of course the present spin dimer theory completely
misses the fact that the excitations really consist of a free two spinon
continuum since it starts from local dimer excitations which could be
interpreted as two spinon bound states.

\subsection{Excitations for weakly coupled dimerized chains, the transverse
dispersion problem} The results of the last section give confidence that the
basic properties of magnetic excitations in dimerized spin chains are correctly
described by the dimer RPA- theory. The advantage of this approach, aside from
its simplicity lies in the fact that it can easily be extended to include
inter- chain coupling. These couplings may lead to transverse dispersion with
$\vec{q}$$\perp\vec{b}^*$, i.e. a dependence of excitation energy on q$_x$ in
addition to the intra- chain dispersion or dependence on q$_y$. As mentioned
previously the q$_x$- dispersion may give important clues about the underlying
CO- structure.\\ First we consider the zero- field case: Then again
Eq.(\ref{RPA}) factorizes into x,y- polarisations but now with sublattice-
exchange terms for each polarisation given by ($\alpha$ =x,y)

\begin{eqnarray} J^\alpha_{AA}(\vec{q})&=&J^\alpha_{BB}(\vec{q})\equiv
J^\alpha_D(\vec{q}) \nonumber\\
J^\alpha_{AB}(\vec{q})&=&J^\alpha_{BA}(\vec{q})^*\equiv J^\alpha_N(\vec{q})
\end{eqnarray}

where J$^\alpha_D$, J$^\alpha_N$ refer to intra- and inter- sublattice
exchange with A,B denoting the two inequivalent {\em dimer} sublattices of the
CO- structures. The four magnetic excitation branches of the planar system of
chains in Fig.1a,b are obtained as solutions of

\begin{equation}
1-u_{\alpha\alpha}(\omega)[J^\alpha_D(\vec{q})\pm|J^\alpha_N(\vec{q})|]=0
\label{Pole} \end{equation}

The choice of $\pm$ in this equation determines the frequency of the acoustical
(A) or optical (O) mode with respect to the two sublattices.\\

{\em In-line chain structure:}\\ We first discuss the in-line CO structure of
Fig.1a. It has exchange Fourier components

\begin{eqnarray} J^\alpha_D(\vec{q})&=&\frac{1}{2}J_2^\alpha\cos2q_y\nonumber\\
J^\alpha_N(\vec{q})&=&-J'^\alpha_d\sin q_y\sin\frac{1}{2}(q_x+q_y)
\end{eqnarray}

Together with Eq.(\ref{Single}) the above equations lead to the four mode
dispersions

\begin{eqnarray} \omega_{x\pm}^2(\vec{q})&=&\frac{1}{2}(J_y+J_z)
[\frac{1}{2}(J_y+J_z)(1+\delta)^2\nonumber\\
&&-J_x(1-\delta^2)\cos2q_y]\nonumber\\ &&\pm(J_y+J_z)J'^x_d(1+\delta)\sin
q_y\sin\frac{1}{2}(q_x+q_y)\\ \omega_{y\pm}^2(\vec{q})&=&\frac{1}{2}(J_x+J_z)
[\frac{1}{2}(J_x+J_z)(1+\delta)^2\nonumber\\
&&-J_y(1-\delta^2)\cos2q_y]\nonumber\\ &&\pm(J_x+J_z)J'^y_d(1+\delta)\sin
q_y\sin\frac{1}{2}(q_x+q_y) \label{CDisp}\nonumber \end{eqnarray}

An interesting aspect of this equation is that for q$_y$= 0, $\pi$ where the
excitation energy is close to the spin gap $\Delta_s$, there is {\em no}
transverse dispersion for the in-line CO model along the lines (q$_x$,0) and
(q$_x$,$\pi$). The dispersion of modes for the present case is shown in Fig.2,
unfolded in the (q$_x$,q$_y$)- plane. The inter-chain coupling J'$_d$ has its
largest effect at the maximum mode energy along (q$_x$,$\frac{\pi}{2}$) where
it causes an additional acoustic(A)- optic(O) mode splitting connected with the
$\pm$ in the above equation and in addition it leads to a q$_x$- dispersion. On
the other hand when $\vec{q}$ = (q$_x$,0) or (q$_x$,$\pi$) J'$_d$ has no effect
and the observed mode splitting in Fig.3 is dispersionless, it is not of A-O
type but has pure anisotropy character as in the single chain case of
Eq.(\ref{Anis}).

{\em zig-zag chain structure:}\\ In Sec.IV.A it was noted that for a single
chain this model is equivalent to the in-line structure. However it can be seen
from Figs.1a,b that the inter- chain coupling is different in the two models.
For the in-line structure a given dimer is symmetrically coupled with $\pm$
J'$_d$ to four dimers on two neighboring chains whereas in the zig-zag model
the coupling is asymetric with strength -J'$_d$, $\frac{1}{2}$J'. This leads
now to Fourier components for the exchange given by

\begin{eqnarray} J^\alpha_D(\vec{q})&=&\frac{1}{2}J^\alpha_2\cos2q_y\\
J^\alpha_N(\vec{q})&=&\frac{1}{2}J'^\alpha\cos(\frac{3}{2}q_y+\frac{1}{2}q_x)
-J'^\alpha_d\cos(\frac{3}{2}q_y-\frac{1}{2}q_x)\nonumber \end{eqnarray}

Using Eq.(\ref{Pole}) we obtain the explicit solutions (with the formal
replacement J$_d\rightarrow$ J)

\begin{eqnarray} &&\omega_{x\pm}^2(\vec{q})=\nonumber\\ &&\frac{1}{2}(J_y+J_z)
[\frac{1}{2}(J_y+J_z)(1+\delta)^2-J_x(1-\delta^2)\cos2q_y]\nonumber\\
&&\pm[(J_y+J_z)[J'^x_d\cos(\frac{3}{2}q_y-\frac{1}{2}q_x)
-\frac{1}{2}J'^x\cos(\frac{3}{2}q_y+\frac{1}{2}q_x)]\nonumber\\
&&\omega_{y\pm}^2(\vec{q})=\\ &&\frac{1}{2}(J_x+J_z)
[\frac{1}{2}(J_x+J_z)(1+\delta)^2-J_y(1-\delta^2)\cos2q_y]\nonumber\\
&&\pm[(J_x+J_z)[J'^y_d\cos(\frac{3}{2}q_y-\frac{1}{2}q_x)
-\frac{1}{2}J'^y\cos(\frac{3}{2}q_y+\frac{1}{2}q_x)] \label{ZDisp}\nonumber
\end{eqnarray}

While the intra-chain part in this expression is the same as in the in-line
model of Eq.(\ref{Anis}) the second part leading to the transverse q$_x$-
dispersion is completely different. For example taking q$_y$= $\pi$ we obtain

\begin{eqnarray} &&\omega_{x-}^2(q_x,\pi)-\omega_{x+}^2(q_x,\pi)=\nonumber\\
&&2(J'^x_d+\frac{1}{2}J'^x)(J_y+J_z)\sin\frac{1}{2}q_x\nonumber\\
&&\omega_{y-}^2(q_x,\pi)-\omega_{y+}^2(q_x,\pi)=\\
&&2(J'^y_d+\frac{1}{2}J'^y)(J_x+J_z)\sin\frac{1}{2}q_x \label{xdisp}\nonumber
\end{eqnarray}

This shows that in addition to the anisotropy induced x,y- mode splitting each
of them shows a further splitting into A,O ($\pm$)- modes which has dispersion:
it vanishes at q$_x$=0 and is at maximum for q$_x$=$\pi$. This situation is
clearly illustrated in Fig.3. Whether this dispersion is visible in the
experiment depends on how large it is against the pure anisotropy splitting
caused by the intra- chain exchange. In principle both are present and Fig.3b
shows two typical possibilities. The q$_x$ {\em dispersive A-O} splitting which
is absent for q$_y$= 0, $\pi$ in the in-line case therefore in principle offers
a possibility to distinguish between both models.

Finally we discuss the intensity variation of q$_y$= 0, $\pi$ spin gap modes as
function of total momentum transfer $\kappa$= $\vec{q}+\vec{\tau}$
($\vec{q}\in$ 1.BZ) mentioned in Ref.~\onlinecite{Yosi}. It was observed that
the intensity of the $\hbar\omega$= 10 meV excitation exhibits unexpected
variation in $\tau_x$ with period h=3 where $\vec{\tau}$= (2$\pi$h,2$\pi$k,0)
is a reciprocal lattice vector in the ab- plane. For a strictly 1D system the
intensity should rather be constant and therefore this variaton possibly points
to a more 2D character of magnetic excitations. We now analyze the intensities
in the dimer RPA model for that structure. For simplicity we neglect the
additional splitting of modes caused by xy- exchange anisotropy, i.e. we assume
J$_x$=J$_y$. Then the observed splitting along q$_x$ is entirely an A-O
splitting due to the fact that the {\em dimer} lattice consists of two
sublattices. In this case the intensities may be obtained from the dynamical
susceptibilities \cite{Mack} decomposed according to

\begin{eqnarray} \tensor{\chi}(\vec{q}+\vec{\tau},\omega)&=&
\frac{1}{2}(1+\cos\Phi)\tensor{\chi}_A(\vec{q},\omega)+\\
&&\frac{1}{2}(1-\cos\Phi)\tensor{\chi}_O(\vec{q},\omega) \label{AO}\nonumber
\end{eqnarray}

where $\phi$=$\vec{\tau}\cdot\vec{\rho}$ and $\vec{\rho}$=($\frac{1}{2}$,
-$\frac{1}{2}$) is the vector joining the two {\em dimer} sublattices in units
of a,b respectively (Fig.1b) which leads to $\Phi$=h$\pi$-k$\pi$. From the
imaginary part of $\chi(\vec{q},\omega)$ the A,O intensities are obtained as
($\vec{q}\in$ 1.BZ)

\begin{equation}
I_{A,O}(\vec{q}+\vec{\tau})=\frac{\Delta}{2\omega_{A,O}(\vec{q})}
[1\pm\cos(h\pi-k\pi)] \end{equation}

where $\pm$ corresponds to A,O respectively. In the experiments \cite{Yosi} one
has $\vec{q}$=(q$_x$,$\pi$) and $\vec{\tau}$ given by (2$\pi$h,0). Neglecting
the small A-O splitting, i.e. setting $\omega_{A,O}(\vec{q})\simeq\Delta_s$ one
then has

\begin{equation} I_{A,O}(\vec{\tau})=\frac{\Delta}{2\Delta_s}(1\pm\cos h\pi)
\label{Intens} \end{equation}

Two points are worth noting: The period of the intensity is given by h=2 and
not h=3. The intensity maxima of slightly split A,O- modes are shifted by one
half period (h=1). Experimentally the intensity at an energy transfer
$\hbar\omega$= 10 meV was measured as function of h. Since this energy is just
in between upper (O) and lower (A) mode and both have a line width considerably
higher than their splitting the measured intensity is then the {\em average} of
A and O mode intensity. According to Eq.(\ref{Intens}) however the average is a
constant independent of h, irrespective of the period of individual A,O
intensities. We conclude that the zig-zag CO structure, at least in the dimer
RPA model for weakly coupled zig-zag chains, does neither explain the observed
intensity variation nor its period.

{\em field dependence of excitations:}\\ Investigation of the field dependence
of magnetic excitations may give further information on the nature of the spin
gap and its observed splitting and transverse dispersion. The field dependence
may also be calculated from the basic Eq.(\ref{RPA}) where it enters through
the local dimer susceptibility Eq.(\ref{HSingle}). Due to breaking of time
reversal symmetry there is now a nondiagonal term u$_{xy}$= -u$_{yx}^*$ and
both polarisations couple to each of the $\Delta$(h), $\Delta'$(h) local dimer
transitions. The poles of Eq.(\ref{RPA}) are then given by

\begin{eqnarray}
&&1-[u_{xx}(\omega)J_\pm^x(\vec{q})+u_{yy}(\omega)J_\pm^y(\vec{q})]\nonumber\\
&&+[u_{xx}(\omega)u_{yy}(\omega)-|u_{xy}(\omega)|^2]
J_\pm^x(\vec{q})J_\pm^y(\vec{q})=0 \label{HPole} \end{eqnarray}

Here J$_\pm^\alpha(\vec{q})$= J$_D^\alpha(\vec{q})\pm$J$_N^\alpha(\vec{q})$ and
$\pm$ has to be taken synchronously at all positions. After straightforward but
lengthy algebra the solution of this equation leads to the field dependent
dispersions for the magnetic excitation branches $\omega_\kappa(\vec{q},h)$
($\kappa$ =1-4):

\begin{eqnarray} \omega_\kappa^2(\vec{q})&=&\frac{1}{2}B_\sigma\pm
\frac{1}{2}[B_\sigma^2(\vec{q},h)-4C_\sigma(\vec{q},h)]^\frac{1}{2} \nonumber\\
B_\pm(\vec{q},h)&=&\Delta^2(h)+\Delta'^2(h)\nonumber\\
&&-2[\Delta(h)J_\pm^x(\vec{q})+\Delta'(h)J_\pm^y(\vec{q})]u^2\nonumber\\
&&-2[\Delta(h)J_\pm^y(\vec{q})+\Delta'(h)J_\pm^x(\vec{q})]v^2 \nonumber\\
C_\pm(\vec{q},h)&=&\Delta^2(h)\Delta'^2(h)-\\
&&-2[\Delta(h)J_\pm^y(\vec{q})+\Delta'(h)J_\pm^x(\vec{q})]\Delta(h)\Delta'(h)u^2
\nonumber\\
&&-2[\Delta(h)J_\pm^x(\vec{q})+\Delta'(h)J_\pm^y(\vec{q})]\Delta(h)\Delta'(h)v^2
\nonumber\\
&&+4J_\pm^x(\vec{q})J_\pm^y(\vec{q})[(\Delta^2(h)+\Delta'^2(h))u^2v^2+\nonumber\\
&&\Delta(h)\Delta'(h)(u^4+v^4)]\label{HSolu} \nonumber \end{eqnarray}

Here $\sigma=\pm$ and $\kappa=(\pm,\sigma)$ corresponds to any of the four
possible combinations of $\pm$- signs in the last equation. In the quantities
B$_\pm$ and C$_\pm$ the $\pm$ signs always have to be taken simultaneously.
With u(h) and v(h) given by Eq.(\ref{Trans}) and $\Delta(h)$, $\Delta'(h)$ by
Eq.(\ref{Delta}) the above expressions represent the complete solution for the
field dependent dispersion of magnetic excitations in the anisotropic coupled
dimer system. These equations can be applied to the CO structures of Figs.
1,6a. The specific CO determines only the exchange functions
J$_\pm^\alpha(\vec{q})$. For zero field this equation reduces to the previously
studied solutions of Eq.(\ref{Pole}). Figure 4 shows the field dependence of
$\vec{q}$=(0,$\pi$) modes, i.e. the spin gap modes vs. external field for the
zig-zag CO in the two limiting cases corresponding to Fig.3b. One obtains a
quasi- linear Zeeman splitting of $\vec{q}$=(0,$\pi$) modes in the small
anisotropy (2) case and almost field independent modes for large anisotropy
(1). The gap will close only at a very high field which is expected since the
zero field spin gap of $\Delta_s$ =10 meV is quite large. In this model it was
assumed that the dimerization $\delta$ itself shows little field dependence
since it should be a direct consequence of the lattice superstructure induced
by the CO.\\ As in the zero field case the susceptibility
$\tensor{\chi}(\vec{q},\omega)$ in Eq.(\ref{RPA}) may also be used to calculate
the intensity of the four $\omega_\kappa(\vec{q},h)$ excitation branches. They
are given by the imaginary part $\pi^{-1}\chi^\perp(\vec{q},\omega)$" where
$\chi^\perp$=$\frac{1}{2}(\chi_{xx}+\chi_{yy})$. One obtains delta- function
contributions of the type $Z_\kappa\delta(\omega-\omega_\kappa(\vec{q}))$. The
intensity of each mode $\omega_\kappa(\vec{q},h)$ can be obtained from
Eq.(\ref{HPole}) as

\begin{eqnarray} &&Z_\kappa(\vec{q},h)=\pm(\frac{\Delta+\Delta'}{2\omega_\pm})
(\frac{\omega_\pm^2-\frac{1}{2}\gamma\frac{J_\pm^x(\vec{q})+J_\pm^y(\vec{q})}
{\Delta+\Delta'}}{\omega_+^2-\omega_-^2})\\
&&\gamma=4[(\Delta^2(h)+\Delta'^2(h))u^2v^2+\Delta(h)\Delta'(h)(u^4+v^4)]
\label{Weight}\nonumber \end{eqnarray}

In the isotropic case with J$_\pm^x$= J$_\pm^y$ and $\Delta(h)=\Delta'(h)$ this
reduces to a simple formula for the two (A,O) modes ($\sigma =\pm$):
Z$_\sigma$($\vec{q}$,h)= $\Delta(h)$/$\omega_\sigma(\vec{q},h)$ which
corresponds to the prefactor of the zero field result in Eq.(\ref{Intens}), the
variaton with $\vec{\tau}$ is supressed here. Within the field range of Fig.4
there is only a few per cent change of the corresponding mode intensity.

\subsection{Excitations in strongly coupled chains}

One reason for focusing on 1D chain models for the magnetic excitations of \Va
was the observation of quasi- 1D temperature dependence of the susceptibility
in the MV phase above T$_c$. This was attributed to d-electrons localised in
the molecular bonding orbitals of each V-V rung having strong exchange J along
the ladder and weak exchange J' between them. This picture was qualitatively
supported by by LDA- calculations \cite{Smol} mapped on an effective 3d- tight
binding (TB) model which lead to very small hopping matrix elements t'$\ll$t
suggesting that J'=$\frac{4t'^2}{U}$ $\ll$ J=$\frac{4t^2}{U}$ in a simple
superexchange picture. However a recent LDA+U analysis \cite{Yar} with a
mapping to an extended TB- model including both V3d and O2p orbitals has
seriously questioned this picture for the low temperature CO phases. In this
calculation the mapping of LDA+U total energies of various CO and spin
polarized states to that of a corresponding Heisenberg model enables one to
calculate realistic values for the most important exchange constants. It turns
out that in the CO phase J' is only about a factor of two smaller than J$_d$
and this "diagonal" ladder exchange is even bigger than the exchange J along
the leg of the ladder. Furthermore surprisingly even the J'$_d$ exchange
constant is not negligible and both J' and J'$_d$ are ferromagnetic. For a
realistic value of U= 3eV the exchange constants have values as given in the
caption of Fig.(5). The reason for the large J' in the CO structure as compared
to the homogeneous MV state lies in the change of pd- hybridisation due to the
shift of 3d- levels on inequivalent V- atoms \cite{Yar}. If this LDA+U result
for the exchange corresponds to the real situation then CO \Va is magnetically
more like a 2D system with strong AF coupling along the ladder diagonals and
legs and almost equally strong FM coupling between the ladders. Such a model is
very different in principle from the 1D models discussed sofar. We now also
investigate its magnetic excitations and origin of the spin gap which is
different from the dimerization mechanism in this model. This is also partly
motivated by the fact that according to Ref.~\onlinecite{Smaalen} there is
indeed no intra-chain dimerisation in the low temperature structure as assumed
in the previous models. Naturally the dimer approach of previous sections is
not possible for the zig-zag CO structure with its very large interchain
coupling J'. On the other hand there is no problem for the in-line structure
since J' is inactive in this case and even the appreciable J'$_d$ obtained from
LDA+U does not affect the dispersion very much since it is effective only at
the maximum energy and does not influence the spin gap as shown in Fig.2. For
the zig-zag CO instead we now start from a broken symmetry ground state with a
spin configuration as indicated in Fig.(1b) which has the lowest ground state
energy E= $\frac{J'}{8}-\frac{J'_d}{4}-\frac{J_d}{4}$. Of course this approach
does not describe the real ground state of \Va which is nonmagnetic,
nevertheless the excitation spectrum can be expected to have realistic
features. The spin state consists of four magnetic sublattices A$\uparrow$,
B$\uparrow$, A$\downarrow$, B$\downarrow$. The molecular field for the
sublattices $\lambda$ is given by
$\Delta(\uparrow$)=-$\Delta(\downarrow)\equiv\Delta$ with

\begin{equation} \Delta=\langle S\rangle[J'^z-2J'^z_d-2J^z_d] \end{equation}

As in the previous models we include anisotropies in the largest exchange
J$_d^\alpha$ although its magnitude has not yet been calculated in LDA+U which
was applied without spin- orbit coupling \cite{Yar}. Without loss of generality
we assume that the spins are oriented along the c-axis, i.e.
J$_d^z>$J$_d^{x,y}$. Furthermore $\langle S\rangle$ is the saturation moment
equal to 1/2 at T=0. The RPA equation for the spin wave (SW) modes is formally
the same as Eq.(\ref{RPA}) but the dynamical variables are now the individual
spins and not the dimer excitations. Therefore instead of Eq.(\ref{Single}) for
the dynamical suscepibilities we have now u$_{xx}$($\omega$)=
u$_{yy}$($\omega$)$\equiv$ u($\omega$) and u$_{yx}$($\omega$)=
u$_{xy}$($\omega$)$^*\equiv$ v($\omega$) with

\begin{equation} u(\omega)=\frac{\langle S\rangle_\lambda\Delta_\lambda}
	       {\Delta^2_\lambda-\omega^2}\;\;\;\;\;
v(\omega)=i\frac{\langle S\rangle_\lambda\omega}
	       {\Delta^2_\lambda-\omega^2}
\end{equation}

where $\langle S\rangle_\lambda$=$\pm\langle S\rangle$ and
$\Delta_\lambda$=$\pm\Delta$ for $\lambda=\uparrow,\downarrow$ sublattices.
Furthermore the exchange Fourier transforms $\tensor{J}^\alpha_{D,N}(\vec{q})$
($\alpha$ =x,y,z) are now tensors defined in the original spin lattice
($\uparrow,\downarrow$ sublattices) instead of the dimer covering lattice. The
various components connecting the four sublattices can be read off from
Fig.(1b), e.g. J$_{D\uparrow\uparrow}$ =J'($\vec{q})$ etc. with

\begin{eqnarray} J'(\vec{q})&=&J'\exp i\frac{1}{2}(\frac{1}{3}q_x-q_y)
\nonumber\\ J'_d(\vec{q})&=&2J'_d\cos\frac{1}{2}(q_x+q_y)\\
J_d(\vec{q})&=&2J_d\exp(-\frac{1}{3}q_x)[\cos q_y-i\delta\sin q_y]\nonumber
\end{eqnarray}

Here $\delta\ll1$ is the dimerization of the zig-zag chain which may exist due
to the low symmetry of the corresponding CO structrue. After some algebra the
complete RPA spin wave solution of Eq.(\ref{RPA}) applied to the present case
consists of four branches ($\kappa$ =1-4) which are given by
$\hat{\omega}_\kappa$= $\omega_\kappa/\langle S\rangle$ with

\begin{eqnarray} \hat{\omega}^2_\kappa(\vec{q})&=&(c'_1\hat{\pm}d'_1)
\nonumber\\ &&\pm\{(c'_1\hat{\pm}d'_1)^2-[|c_1|^2+|d_1|^2-|c_2|^2-|d_2|^2
\nonumber\\ &&\hat{\pm}(c_1d_1^*+c_1^*d_1-c_2d_2^*-c_2^*d_2)]\}^\frac{1}{2}
\label{SW}\\ c_1&=&\hat{\Delta}^2 +J'^xJ'^{y*} -(J'^x_dJ'^y_d
+J^x_dJ^{y*}_d)\nonumber\\ c_2&=&-\hat{\Delta}(J'^x +J'^y)- (J^x_dJ'^y_d
+J^y_dJ'^x_d)\nonumber\\ d_1&=&\hat{\Delta}(J'^x_d-J'^y_d) +J'^xJ^{y*}_d
-J'^{y*}J^x_d\nonumber\\ d_2&=&\hat{\Delta}(J^x_d -J^y_d) +J'^xJ'^y_d
-J'^yJ'^x_d\nonumber \end{eqnarray}

Here $\hat{\Delta}$=$\Delta/\langle S\rangle$ and c'$_1$=(c$_1$+c$_1^*$)/2,
d'$_1$=(d$_1$+d$_1^*$)/2 denotes the real part of these functions. Note that
the variable $\vec{q}$ was suppressed in J'($\vec{q}$), J'$_d$($\vec{q}$) and
J$_d$($\vec{q}$) in the above expressions for simplicity. The signs $\hat{\pm}$
with a hat have to be taken simultaneously with upper or lower value wherever
they appear thus leading to four spin wave branches. Again it is useful to
consider the solutions for the single chain case only, i.e. setting
J'=J'$_d$$\equiv$0. Then Eq.(\ref{SW}) reduces to
$\hat{\omega}_\kappa^2$=c'$_1\pm|d_2|$ or

\begin{eqnarray} \hat{\omega}^2_\pm(\vec{q})&=& [\hat{\Delta}\hat{\pm}2J^x_d
\gamma_{\vec{q}}] [\hat{\Delta}\hat{\mp}2J^y_d \gamma_{\vec{q}}]\\
\gamma_{\vec{q}}&=&(\cos^2q_y+\delta^2\sin^2q_y)^\frac{1}{2}\nonumber
\end{eqnarray}

Each branch is twofold degenerate since there is no A-O splitting without
inter- chain coupling. At zero wave vector one has for J$_d>$0, using
2$\langle S\rangle$ =1:

\begin{eqnarray}
\omega_+(0)&=&[J^z_d-J^x_d]^\frac{1}{2}[J^z_d+J^y_d]^\frac{1}{2}\simeq
(2j_d^x\bar{J}_d)^\frac{1}{2}\nonumber\\
\omega_-(0)&=&[J^z_d+J^x_d]^\frac{1}{2}[J^z_d-J^y_d]^\frac{1}{2}\simeq
(2j_d^y\bar{J}_d)^\frac{1}{2} \label{SAgap} \end{eqnarray}

with j$_d^{x,y}$= J$_d^z$- J$_d^{x,y}$ and
$\bar{J}_d$=(J$_d^x$+J$_d^y$+J$_d^z$)/3 denoting the exchange anisotropy and
average respectively.

This shows that in the SW approximation the spin gap
$\Delta_s^\alpha$=(2j$_d^\alpha\bar{J}_d)^\frac{1}{2}$ is entirely an
anisotropy gap and independent of CO induced dimerization. Indeed in the
isotropic case $\omega(\vec{q})$=2$\langle S\rangle$
J$_d(1-\delta^2)^\frac{1}{2}$$\sin q_y$. The dimerization does not remove the
gapless excitations but only changes slightly the spinwave velocity. Therefore
not surprisingly for isolated {\em dimerized} isotropic HAF chains the SW RPA
approximation is qualitatively incorrect and the dimer RPA approach of previous
sections should be used. However when interchain couplings J', J'$_d$ become
larger than J$_d\delta$ the situation is reversed and the SW approximation of
Eq.(\ref{SW}) is a better starting point for the essentially 2D magnetic
system. This is certainly the case when one uses the exchange parameters
obtained from LDA+U where due to $|J'/J_d|\simeq$ 0.5 J' is much larger than
J$\delta$ with a $\delta$=0.034 estimated from the spin gap in the dimer
model.\\ The SW excitation branches for the 2D exchange model as obtained from
the LDA+U parameters are shown in Fig.(5). The size of the anisotropy which is
not determined by LDA+U is fixed by the size of the spin gap of $\simeq$ 10 meV
as suggested by Eq.(\ref{SAgap}). For zero anisotropy one would get a Goldstone
mode (A- branch) also for the 2D model. Thus the anisotropy and not the CO
induced dimerization of the J$_d$ exchange is the origin of the spin gap in
this 2D model. Fig.(5) exhibits a pair of two strongly split A,O modes where
the splitting is approximately given by $[J'(J'-2J_d)]^\frac{1}{2}$. Both modes
show a small additional splitting caused by the xy- exchange anisotropy of
J$_d^\alpha$; if it vanishes the A and O modes are twofold degenerate
throughout the BZ. This fact is well known already from the simple (two
magnetic sublattice) AF where only A modes exist and the degeneracy can simply
be understood as a result of the downfolding into the AF BZ. The A- modes are
nearly dispersionless along q$_x$ because the effect of the AF J$_d$ and the FM
J'$_d$ nearly cancel along this direction. The behaviour of the A-modes around
the (0,$\pi$)- point with their splitting increasing towards ($\pi$,$\pi$) is
qualitatively very similar to the two modes in the dimer model with weakly
coupled chains (Figs. 2,3). Note however that the role of A-O splitting and
anisotropy splitting are reversed. It is the latter which now leads to the
dispersion of the small A-mode splitting whereas the A-O splitting is now much
larger. The existence of the high energy split- off O- branch is an essential
prediction of this 2D model and could  be tested directly experimentally. The
flat part of the O- branch lies at about an energy given by $\omega_O\simeq
[J'(J'-2J_d)]^\frac{1}{2}\simeq$ 30 meV roughly twice the maximum energy
investigated so far \cite{Yosi}.

\subsection{Excitations in Ladder and partly mixed valent structures}

Recently a model for the lattice distortion caused by the charge ordering below
T$_c$ has been proposed based on low temperatue x-ray scattering
\cite{Smaalen}. According to this model only every second ladder in Fig.1c is
distorted {\em perpendicular} to the chain ($\vec{b}$-) direction with a period
of 2a along the ($\vec{a}$-) direction. It is not immediately obvious which CO
structure is compatible with this distortion pattern but the simple in-line and
zig-zag model cannot easily be reconciled with the curious fact that every
second V-ladder of the Trellis lattice is undistorted. Two possible models
which incorporate this fact have been proposed and are shown in Fig.6: (i) CO
structure of the spin ladder type \cite{Smaalen1,Regnault} where {\em two}
V(3d) electrons occupy every second rung and the other rungs have no V(3d)
electrons. This ladder model is therefore completely different from the chain
models which have only {\em one} 3d- electron for every rung. It seems rather
surprising that the ladder structure should be realized since LDA+U
calculations \cite{Yar} indicate that it has a much higher total energy than
the chain structures. (ii) structure with alternating CO zig-zag chains and
disordered chains \cite{Smaalen1}. Here half of the ladders have zig-zag CO
like in Fig.1b but the other half remains in the disordered mixed valent state.
This is again a structure with only one d- electron per rung on the average and
therefore it will have a total energy not too different from the CO structure
of Fig.1b.\\ We discuss first the magnetic excitations in the ladder structure
in qualitative terms assuming an AF superexchange $\tilde{J}^\alpha$ of spins
within the ladder rungs via the intervening oxygen. In this structure there is
no connection between the spin gap and the doubling of periodicity along b
since it appears already in the undimerized (equidistant) ladder. Essentially
one deals with single ladder excitations in this case because the magnetic
S=$\frac{1}{2}$ V$^{4+}$ ladders are separated by nonmagnetic S=0 (V$^{5+}$)
ladders. There is another important difference to the chain structures: The
doubling of the periodicity of the lattice (period 2b) does not show up in the
spin ladder which still has periodicity b as seen in Fig.1c., this has drastic
consequences for the excitations. In the present case the local dimer
susceptibility is determined by the states of the rung- dimer with interaction
$\tilde{J}^\alpha$ (Fig. 1c) and it may be obtained by replacing
$\Delta\rightarrow$ $\frac{1}{2}(\tilde{J}^y+\tilde{J}^z)$ and $\Delta
'\rightarrow$ $\frac{1}{2}(\tilde{J}^x+\tilde{J}^z)$ in Eq.(\ref{Single}).
Using Eq.(\ref{Pole}) which also holds for the present case we obtain for the
two excitation branches (there is no A-O splitting in this case):

\begin{eqnarray} \omega_x^2(\vec{q})&=&\frac{1}{2}(\tilde{J}_y+\tilde{J}_z)
[\frac{1}{2}(\tilde{J}_y+\tilde{J}_z)-2J_e^x\cos q_y]\nonumber\\
\omega_y^2(\vec{q})&=&\frac{1}{2}(\tilde{J}_x+\tilde{J}_z)
[\frac{1}{2}(\tilde{J}_x+\tilde{J}_z)-2J_e^y\cos q_y] \end{eqnarray}

The most important aspect of these excitations is the doubling of the
periodicity 2$\pi$ along the ladder instead of $\pi$ for the chain models. In
the latter the points q$_y$=0 and q$_y$=$\pi$ are degenerate and their
excitation energy is at (or close) to the minimum, i.e. equal to the spin gap
(Figs.2,3). On the other hand for the ladder structure assuming
J$_e^\alpha$=J$_d^\alpha$-J$^\alpha <0$ the excitation energy is equal to the
spin gap only for q$_y$=$\pi$, whereas it is at the maximum in the zone center
(q$_y$=0) for AF effective inter-(rung) dimer exchange J$_e$. The possibility
of the spin ladder model can therefore in principle be directly experimentally
investigated. Finally the splitting along the transverse q$_x$ direction will
be constant and determined by the exchange anisotropy as in the in-line chain
case.\\

The more recent proposal \cite{Smaalen1} of an alternating CO/MV low
temperature structure of \Va which is shown in Fig.6b is a very interesting
possibility and deserves a detailed analysis of its magnetic excitation
spectrum. In this structure CO zig-zag chains alternate with ladders in the MV
state along the a- axis. In both we have one d-electron per rung on the average
but in the MV ladders the electron is not localised on one of the two
V-positions of the rung but resonates, i.e. it is in the molecular bonding
state of the two V-atoms. This means there are three types of V-sites with
formal valencies Z=4 or 5 on the CO zig-zag chain and Z=4.5 on the MV ladder.
Whether the existence of three inequivalent V- sites is compatible with NMR
results is not clear. To describe the magnetic excitations in such an
inhomogeneous state we make a drastic but reasonable assumption: Since the
d-electrons residing in the molecular orbitals are spread out over the whole
rung, their spin response will be concentrated around zero total momentum
transfer contrary to the atomic spins on the CO zig-zag chains where the
magnetic scattering intensity varies with the {\em atomic} form factor.
Therefore the contribution of the spins in the molecular orbitals to the
scattering cross section should be negligible at large momentum transfer and
consequently {\em only the atomic} spins residing on the zig-zag chains will be
included in the model exchange Hamiltonian for the strucure of Fig.6b. This
model Hamiltonian has only two (in general anisotropic) exchange constants:
intra-chain J$_d^\alpha$ similar as in the zig-zag model of Fig.1b and
inter-chain coupling J$_l^\alpha$ which connects V$^{4+}$ spins on next nearest
ladders via a superexchange path across the intervening MV (non- magnetic)
ladder. Note that in principle the lattice distortion in \cite{*} which was
described in the beginning of this section implies a dimerization J$_{1,2}$=
J$_d$(1$\pm\delta$) of the intra- chain J$_d^\alpha$ {\em along the}
$\vec{a}$-{\em direction}. This means that the intra-chain exchange within a
given zig-zag chain is J$_d$(1+$\delta$) and J$_d$(1-$\delta$) on two adjacent
zig-zag chains separated by a MV ladder(Fig.6b). This type of dimerization
therefore leads to a doubling of the unit cell of the exchange Hamiltonian
along $\vec{a}$ whose consequences we will discuss later. Note that it does
{\em not} lead to an alternating exchange within a given zig-zag chain and
hence {\em not} to a dimerization gap in the excitations of the isolated chain.
For this reason it is possible to start from a N\'{e}el ground state and use
the spin wave approximation even though we consider now weakly coupled zig-zag
chains. Formally the SW- calculation is very similar to that performed in
section IV.C and we do not repeat the details. The Fourier components of the
exchange entering the dynamical susceptibility $\chi(\vec{q},\omega)$ may be
read off from the structure in Fig. 6b and are given by

\begin{eqnarray} J_d^\alpha(\vec{q})&=& 2J_d^\alpha \cos
q_y\exp(\frac{i}{3}q_x)\nonumber\\ J_l^\alpha(\vec{q})&=& 2J_l^\alpha \cos
q_y\exp(\frac{2i}{3}q_x) \end{eqnarray}

The spin wave branches are then again obtained from the poles of
$\chi(\vec{q},\omega)$ which leads to the secular equation
$\det\chi(\vec{q},\omega)$=0. In this equation the dimerization
$\delta$($\parallel\vec{a}$) appears only in \it O\rm($\delta^2$) and if these
terms are neglected for the moment we obtain the four spin wave branches in the
small BZ ($\mid$ q$_x$ $\mid$ $\leq$ $\frac{\pi}{2}$) as

\begin{eqnarray}
\omega_\kappa^2(\vec{q})&=&[(J_d^{z2}-J_d^xJ_d^y)+J_d^xJ_d^y\sin^2q_y] \pm
J_d^z(J_d^y-J_d^x)\cos q_y \nonumber\\ &&\pm J_l(J_d^x+J_d^y)\cos^2q_y\cos q_x
\label{COMV} \end{eqnarray}

Here $\pm$ signs are chosen independently and therefore $\kappa$ =1-4. For the
isolated chain with uniaxial exchange (J$_l$=0, J$_d^x$=J$_d^y$ $\equiv$ J$_d$)
this reduces to a single dispersion

\begin{eqnarray} \omega^2(q_y)=\Delta_a^2+J^2\sin^2q_y \end{eqnarray}

with the Ising anisotropy spin gap $\Delta_a^2$= (J$_d^{z2}$-J$_d^{2}$)
$\simeq$2J$_d^z$(J$_d^z$-J$_d$). In the Heisenberg case  $\Delta_a$=0 and one
obtains the dispersion Eq.(\ref{DCP}) but now with $\alpha_{SW}$=1 which is
smaller than the values of $\alpha$ in the dimer approximation and in the exact
result (see below Eq.(\ref{DCP})). For J$_d^x\neq J_d^y$ there are two
anisotropy gaps $\Delta_a^\pm$.\\ We now focus on the most interesting part of
the transverse dispersion along q$_x$ of the coupled chain system. For q$_y$=0
(or q$_y$=$\pi$, this leads only to an interchange of modes) one obtains in the
{\em large} BZ ($\mid$ q$_x$ $\mid$ $\leq\pi$) {\em two unfolded} modes given
by

\begin{eqnarray} \omega_+(q_x)&=&\Delta_a^+ -\delta^+\cos q_x \nonumber\\
\omega_-(q_x)&=&\Delta_a^- -\delta^-\cos q_x \label{Regnault} \end{eqnarray}

with $\Delta_a^{\pm 2}=2J_d^z(J_d^z-J_d^{x,y})$ and
$\delta^\pm=(J_dJ_l)/\Delta_a^\pm$. In deriving Eq.(\ref{Regnault}) from to
Eq.(\ref{COMV}) we assumed that $\delta^\pm\ll$1. These spin wave dispersions
are equivalent to the recently proposed empirical dispersion
relations\cite{Regnault} obtained by new inelastic neutron scattering results
which had much higher resolution than those performed in
Ref.~\onlinecite{Yosi}. The above formulas provide an excellent fit to the
experimental dispersions as shown by Regnault et al in
Ref.~\onlinecite{Regnault} with parameters obtained there as $\Delta_a^+$=
10.65 meV, $\Delta_a^-$= 8.75 meV, $\delta^+$= 0.4 meV, $\delta^-$= 0.5 meV.
Note that this dispersion proposed by Regnault et al \cite{Regnault} and
derived in the present model has only {\em half} the period in q$_x$ as
compared to the pure zig-zag model of Fig.1b and Eq.(\ref{ZDisp}). Using the
empirical parameters from above and Eq.(\ref{Regnault}) one may completely
determine the relevant parameters in the present theoretical model: J$_d^z$= 38
meV, J$_l$= 0.11 meV for intra- and inter- chain exchange respectively and for
the intra- chain anisotropies we have J$_z$-J$_x$= 1.49 meV and J$_z$-J$_y$= 1
meV. Note that our model result for $\delta^\pm$ below Eq.(\ref{Regnault})
requires $\delta^+$/$\delta^-$= $\Delta^+$/$\Delta^-$. Experimentally
$\delta^+$/$\delta^-$= 0.80 and $\Delta^+$/$\Delta^-$= 0.82 which is close to
expectation. The dispersion obtained from Eq.(\ref{COMV}) (or approximately
from Eq.(\ref{Regnault})) with these parameters is shown in Fig. 7. Finally we
comment on the question of intensities. Since we have two chemical sublattices
separated by $\vec{d}=\frac{2}{3}\vec{a}+\vec{b}$ (Fig.6b). and only
sublattices of opposite spins are coupled in pairs the intensities will be
given by a similar expression as in Eq.(\ref{AO}):

\begin{eqnarray}
I_A(\vec{\tau})\sim\frac{1}{2}(1+\cos(\frac{2\pi}{3}h))&=&\cos^2(\frac{\pi}{3}h)
\nonumber\\
I_O(\vec{\tau})\sim\frac{1}{2}(1-\cos(\frac{2\pi}{3}h))&=&\sin^2(\frac{\pi}{3}h)
\end{eqnarray}

In this case then the period of the intensity variation is correctly given by
h=3 as observed experimentally contrary to the zig-zag model of Fig.1b.\\ Sofar
the dimerisation $\delta$ along $\vec{a}$ leading to J$_d\rightarrow$
(1$\pm\delta$)J$_d$ has been neglected. If included it would lead to an opening
of an additional gap at the points q$_x$=$\pm\frac{\pi}{2}$ of size
2$\delta\Delta_a$ as shown in Fig. 7. Sofar this dimerization gap has not been
identified. It is not clear whether the resolution is still too small or
whether the inter-chain exchange dimerization J$_d$(1$\pm\delta$) of next
neighbor zig-zag chains as shown in Fig. 6b is indeed negligible.

\section{Summary and conclusion}

We have proposed and analyzed a number of spin-excitation models that may be
relevant for inelastic neutron scattering investigations in the CO low
temperature phase of \Va. The most frequently discussed models are based on
in-line or zig-zag chain CO and assume that the exchange coupling along the
chains is much smaller than between the chains. In these models there is a
strong dispersion along the chain axis $\vec{b}^*$ caused by the exchange
coupling along the legs (in-line J) or ladder diagonals (zig-zag J$_d$). Recent
LDA+U calculations \cite{Yar} suggested that both are of the same order of
magnitude. Therefore even in the completely CO zig-zag structure there is a
strong dispersion $\parallel\vec{b}^*$ (q$_y$). The minimum excitation energy,
i.e. the spin gap $\Delta_s$ in this scenario is due to a dimerization
J(1$\pm\delta$) or J$_d$(1$\pm\delta$) of the intra- chain exchange. The
dispersion $\parallel\vec{b}^*$ (q$_x$) on the other hand is comparatively
small. However it shows characteristic differences for the two quasi- 1D
models. Allowing for small anisotropies in the largest exchange J or J$_d$ we
find that (1) the in- line model has a q$_x$- dispersionless splitting of the
spin gap mode for q$_y$= 0,$\pi$ determined by the exchange anisotropy alone.
The inter- chain coupling may only contribute to the q$_x$- dispersion at the
maximum mode energy at q$_y$=$\frac{\pi}{2}$. This is a direct consequence of
the Trellis lattice structure with every second ladder shifted by
$\frac{b}{2}$. (2) in the zig-zag model the splitting of the q$_y$= 0,$\pi$
spin gap modes has both a contribution from exchange anisotropy J$^x_d$-J$^y_d$
and inter- chain coupling J'. Depending whether the former or latter is
stronger one has little or noticeable dispersion along q$_x$ and the role of
anisotropy splitting and A-O splitting are interchanged. It was also shown that
the magnetic field behaviour in the two limiting cases of the zig-zag model
(Fig.4) is different. While for J' appreciably larger than J$_x$-J$_y$ there is
an almost linear Zeeman splitting of spin gap  modes, they are almost field
independent for small fields in the opposite case. In addition we discussed the
zig-zag CO model in the case of strong coupling between the zig-zag chains
because LDA+U calculations predict a surprisingly large inter- chain coupling
J' in this CO- structure. In this case we used a broken symmetry approach to
calculate the spin excitations. In this model the spin gap is a pure anisotropy
gap, furthermore one obtains a split- off optical branch at an energy of 30
meV. The observation of such a mode would be crucial for this model, sofar
there is no experimental evidence that it exists.\\ Furthermore we briefly
discussed the alternative spin ladder model of Fig. 6a with AF coupling in the
ladder rungs. It was found that the dispersion along q$_y$ has twice the
periodicity as compared to the in-line and zig-zag chain models.\\ Perhaps the
most promising model investigated is the mixed CO/MV model with zig-zag chains
separated by disoredered (MV) ladders. This structure model has been proposed
in Ref. \onlinecite{Smaalen1} and we have shown that it leads to spin wave
dispersions exactly as those empirically proposed by Regnault et
al\cite{Regnault} from new inelastic neutrons scattering results. Most
importantly it shows half the period for the dispersion along q$_x$ as compared
to the other models and also has the proper intensity variation. In this model
the spin gap is due to a predominately Ising type exchange anisotropy and the
dimerization perpendicular to the chains has only little effect.\\ All models
discussed account for the basic qualitative properties of the available neutron
scattering results \cite{Yosi}: (1) Sligthly split spin gap modes at
$\Delta_s$= 10 meV with little or no dispersion along q$_x$ and (2) a large
dispersion of magnetic excitations along the chain (q$_y$)- direction. What is
different in the models is the interpretation of the origin of various gaps and
splittings observed, i.e. whether they are due to dimerization, anisotropy,
ladder type or of A-O nature. To make further progress in discriminating
between these models (and possibly others not investigated here) and also to
obtain a more reliable set of exchange parameters for them it is necassary to
have inelastic neutron scattering results in a larger energy and momentum
region and with enhanced resolution and also a more detailed information on the
momentum dependence of the intensity for each individual mode. The
investigation in this paper has given a clear classification of the typical
signatures of the different exchange models one has to look for.

\section*{Acknowledgement} P.T. would like to thank T. Yosihama for information
on Ref. 18 and L.P. Regnault and T. Chatterji for helpful discussions.

\begin{figure} \centerline{\psfig{figure=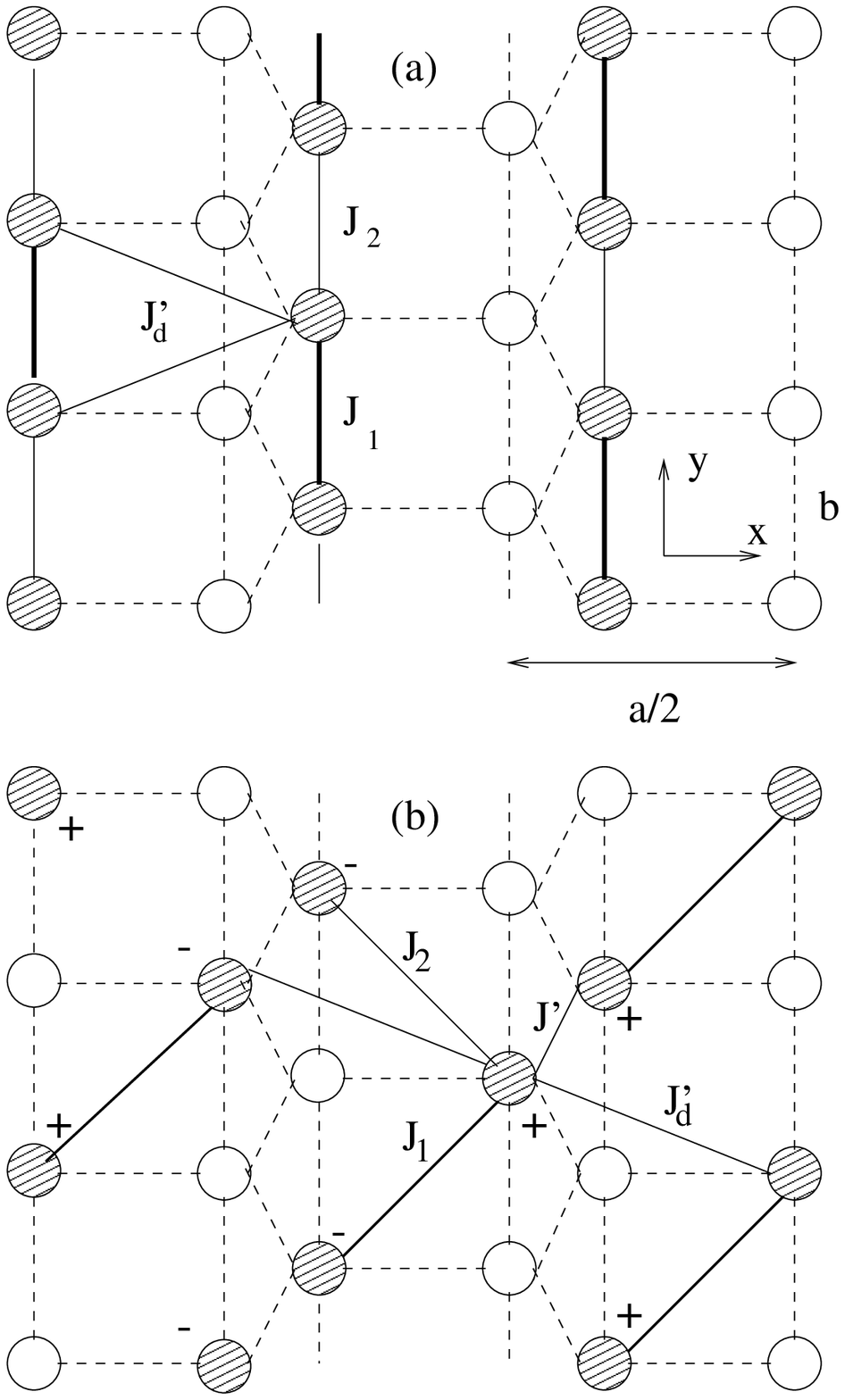,height=12cm,width=6cm}}
\vspace{1cm} \caption{Charge ordered (CO) structures of \Va discussed in the
text. Hatched circles: V$^{4+}$ (S=$\frac{1}{2}$), open circles: V$^{5+}$
(S=0). Oxygen atoms on the legs and rungs of V- ladders are not shown. Thick
lines (J$_1$ or $\tilde{J}$) denote the dimer basis of each model.  (a) in-line
CO with active exchange constants J$_{1,2}$= J(1$\pm\delta$) and J'$_d$;
$\delta$= dimerization strength along b.  (b) zig-zag CO with active exchange
constants J$_{1,2}$= J$_d$(1$\pm\delta$),  J' and J'$_d$. The $\pm$ signs
denote the spin configuration with lowest energy for the LDA+U exchange
parameters. This is only relevant for the 2D spin wave scenario of Sec.IV.C}
\end{figure}

\begin{figure} \centerline{\psfig{figure=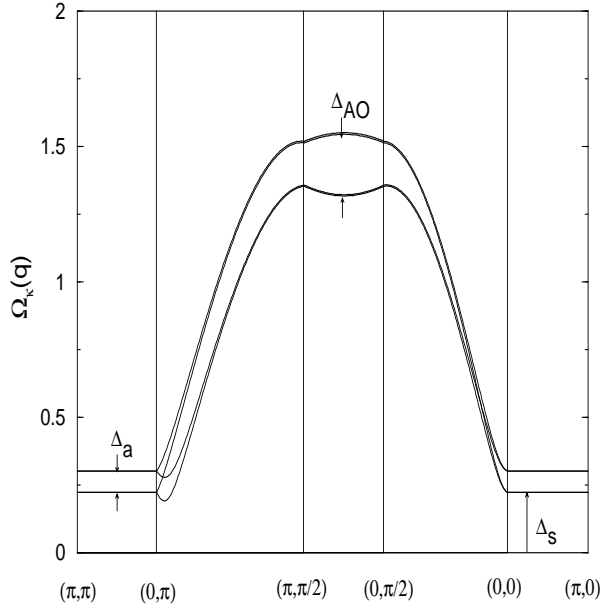,height=8cm,width=8cm}}
\vspace{1cm} \caption{Spin excitations
$\Omega_\kappa(\vec{q})=\omega_\kappa(\vec{q})/J^z$ ($\kappa$=1-4) in the
in-line structure calculated with dimer RPA- theory of Eq.(19). Only J$^\alpha$
($\alpha$=x,y,z)  and J'$_d$ are active exchange constants, the former
determines the large dispersion along $\vec{b}^*$ (q$_y$), the latter the A-O
splitting $\Delta_{AO}$. The spin gap $\Delta_s$ is mainly caused by the
dimerization $\delta$. J'$_d$ has no influence along paths with q$_y$= 0,
$\pi$, therefore in this model $\Delta_a$ is a pure anisotropy splitting
determined by J$_x$-J$_y$ and the two split modes are dispersionless along
q$_x$. In the xy- isotropic case $\Delta_a$ =0. Exchange parameters used are
J$_x$= 38.4 meV,  J$_y$= 37.4 meV, J$_z$= 37.9 meV, J'$_d$= -6 meV and
dimerisation $\delta$= 0.034} \end{figure}

\onecolumn \begin{figure}
\centerline{\psfig{figure=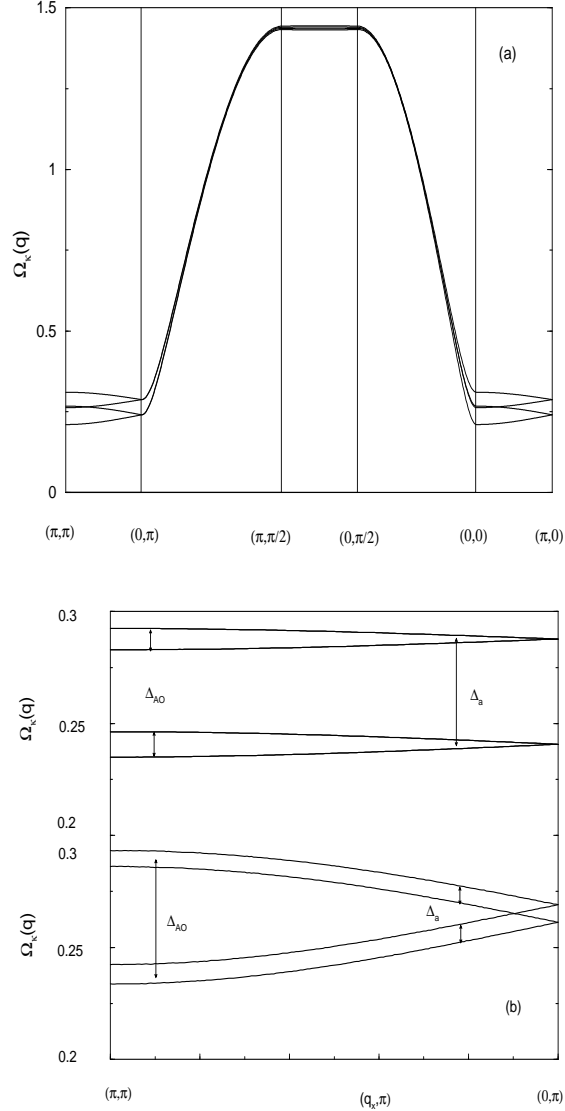,height=16cm,width=8cm}} \vspace{1cm}
\caption{Spin excitations $\Omega_\kappa(\vec{q})=\omega_\kappa(\vec{q})/J_d^z$
($\kappa$=1-4) in the zig-zag CO structure according to Eq.(19) (where the
formal replacement J$_d\rightarrow$ J was made). Only J$_d^\alpha$, J' and
J'$_d$ (Fig.1b) are active exchange constants in this structure.  (a)In
addition to the anisotropy splitting $\Delta_a$ along q$_x$ there is a further
A-O splitting superposed. Which one is more pronounced depends on the relative
size of xy- anisotropy J$_d^x$-J$_d^y$ and the inter- ladder coupling J'.  In
this plot J$_d^x$=38.2 meV, J$_d^y$=37.6 meV, J$_d^z$=37.9 meV, J'= 0.5 meV,
J'$_d$=0 and $\delta$= 0.034 was used.  (b) Enlarged excitation branches
between ($\pi,\pi$) and (0,$\pi$) for two extreme cases. above: large
anisotropy J$_d^x$=38.2 meV, J$_d^y$=37.6 meV, J$_d^z$=37.9 meV and small J'=
0.1 meV. below: small anisotropy J$_d^x$=37.95 meV, J$_d^y$=37.85 meV,
J$_d^z$=37.9 meV and large J'=0.5 meV. Other parameters in both cases as in
(a). It is seen that anisotropy splitting $\Delta_a$ and A-O splitting
$\Delta_{AO}$ interchange roles in the two cases. The situation proposed
previously \cite{Yosi} corresponds more to the lower part. Parameters for the
dispersion in (a) are between these two extreme cases.} \end{figure}

\twocolumn \begin{figure}
\centerline{\psfig{figure=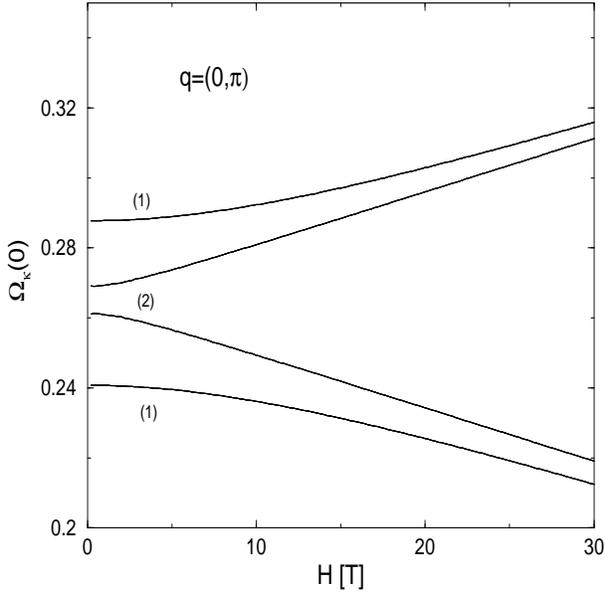,height=8cm,width=8cm}} \vspace{1cm}
\caption{Field dependence of modes at $\vec{q}$=(0,$\pi$) for the zig-zag CO.
Exchange parameters for (1) and (2) are identical to those of the upper and
lower part of Fig.3b respectively.} \end{figure}

\begin{figure} \centerline{\psfig{figure=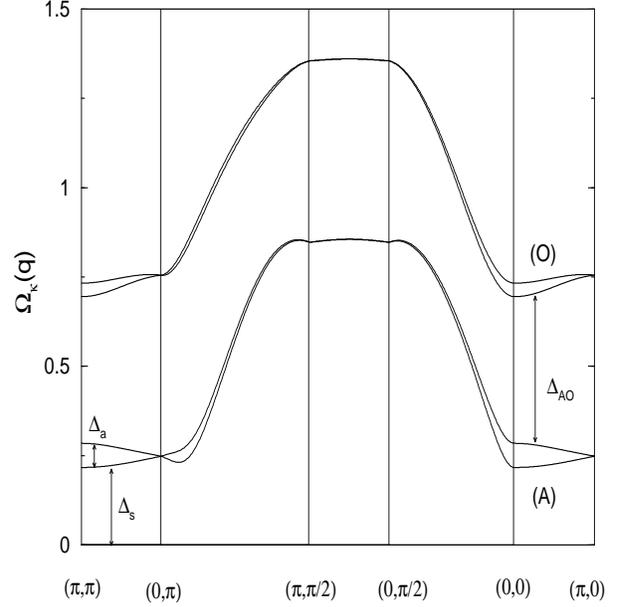,height=8cm,width=8cm}}
\vspace{1cm} \caption{Spin excitations for the 2D exchange scenario described
in Section V. The SW results of Eq.(33) has been used. The exchange parameters
are basically those from LDA+U calculations (U=3 eV) \cite{Yar} for the spin
polarized zig-zag CO structure of Fig.1b, except for a slightly larger J'$_d$
and the additional anisotropies which have been introduced to obtain spin gap
$\Delta_s$ and anisotropy splitting $\Delta_a$. Explicitly we use J$_d^x$=34.5
meV, J$_d^y$=35.2 meV, J$_d^z$=36.2 meV, J'= -17.8 meV, J'$_d$= -6 meV. In the
SW picture the dimerization does not lead to a spin gap and was set to zero.
$\Delta_s$ is mainly determined by the (xy)-z anisotropy and $\Delta_a$ by the
in- plane xy- anisotropy. The large $\Delta_{AO}$- gap is caused by the large
J' in this parameter set.} \end{figure}

\begin{figure} \centerline{\psfig{figure=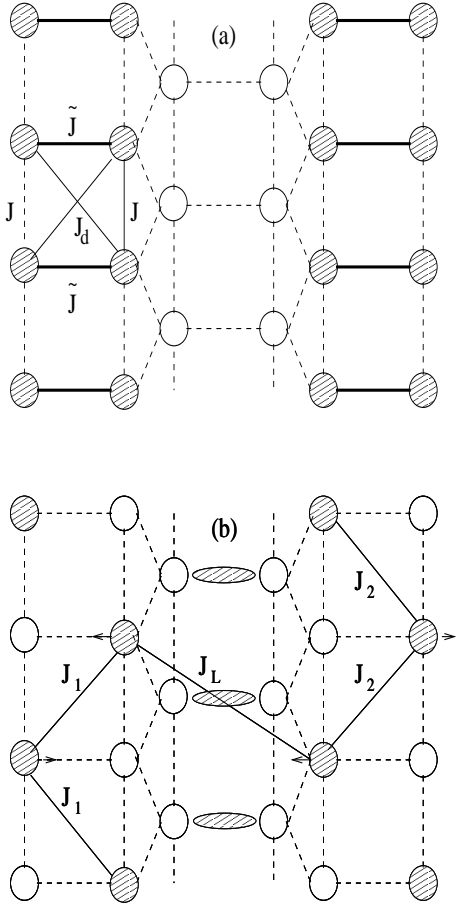,height=12cm,width=6cm}}
\vspace{1cm} \caption{(partly) charge ordered (CO) structures of \Va discussed
in Sec. IV.D. Hatched circles: V$^{4+}$ (S=$\frac{1}{2}$), open circles:
V$^{5+}$ (S=0). Hatched ellipse: V$^{4.5+}$ MV state of V-V rung. Oxygen atoms
on the legs and rungs of V- ladders are not shown.  (a) ladder CO consisting of
isolated S=$\frac{1}{2}$ ladders with active exchange constants J, $\tilde{J}$
and J$_d$.  (b) partly zig-zag CO/MV structure. Intra-chain exchange J$_{1,2}$=
J$_d$(1$\pm\delta$)) is dimerized {\em perpendicular} to the chain under the
assumption of the low temperature distortion pattern (arrows) proposed in Ref.
21. Next neighbor chains are coupled by J$_l$. The hatched ellipse denote one
d-electron in the molecular state of the rung, i.e. the V- atoms have formal
valence 4.5.} \end{figure}

\begin{figure}
\centerline{\psfig{figure=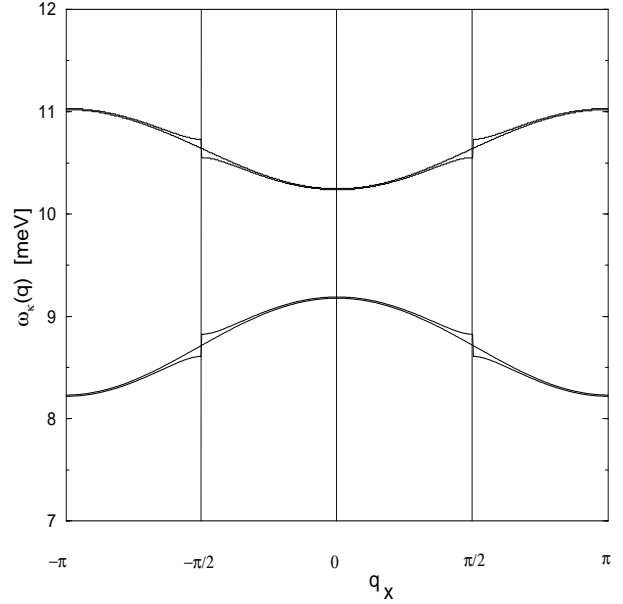,height=8cm,width=8cm,angle=-90}}
\vspace{1cm} \caption{Spin wave dispersion $\omega_\kappa(\vec{q})$ along q$_x$
for q$_y$=0,$\pi$ unfolded in the large BZ corresponding to the undistorted
structure. Gapless curves: Modes for zero dimerisation according to Eq.(37),
gapped curves: modes with {\em perpendicular} dimerization according to Fig.6b
with $\delta$=0.01. This leads to small gaps at the boundary
($\pm\frac{\pi}{2}$) of the small BZ corresponding to the distorted structure.
Exchange parameters and anisotropies have been choosen to comply with the
experimentally determined dispersion by Regnault et al [22] and given in the
text of Sec.IV.D.} \end{figure}

\end{document}